# Effect of phase noise on fringe visibility of LISA telescopes in the presence of tilt and astigmatism aberrations.

Parmest Roy[1,2]

[1]Exchange Student, Sapienza University of Rome, Italy

[2]Undergraduate Student, Department of Physics, Shiv Nadar Institute of Eminence, Delhi NCR

*Abstract: An in-depth analysis of GWs, LISA and its telescopes has been carried out in this paper, especially for the case in which all aberrations, except astigmatism and tilt are optimised. Though it is popular to neglect these two aberration types for their minute effect on the wavefront, a test of this negligence has been carried out in an extreme scenario of high phase noise. The relationship between phase noise and fringe visibility has been analysed, resulting in the discovery of an upper bound for allowed phase noise in LISA.*

## 1. Introduction

The world's first space based gravitational wave detector will be launched in the early 2030s. LISA will consist of three space modules forming the vertices of an equilateral triangle spread across millions of miles in space. [9] Each module will consist of two telescopes, two lasers and a test mass. Gravitational waves will cause a minute change in the length of the sides of the triangle which will be detected by LISA. This will help us collect a giant amount of astrophysical information that has stayed hidden from us till now. However, due to changes in the space environment, the telescopes will be undoubtedly subject to severe additional dynamic aberrations which may bring path errors in the readings of the gravitational wave data. [2] In this paper we will study the effect of those aberration types on LISA that are usually neglected as insignificant.

## 2. Gravitational Waves

Gravitational waves are ripples in spacetime curvature that propagate as waves at the speed of light, generated by certain movements of mass. [7] They are predicted by Einstein's General Theory of Relativity and can be described mathematically as perturbations in the metric tensor of spacetime.

*2.1. Einstein Field Equations:*

The foundation of describing gravitational waves starts with the Einstein field equations:

$$G_{\mu\nu} + \Lambda g_{\mu\nu} = \frac{8\pi G}{c^4} T_{\mu\nu} \qquad (1)$$

where $G_{\mu\nu}$ is the Einstein tensor, $\Lambda$ is the cosmological constant, $g_{\mu\nu}$ is the metric tensor, $G$ is the gravitational constant, $c$ is the speed of light, and $T_{\mu\nu}$ is the stress-energy tensor.

*2.2. Linearized Gravity:*

To describe gravitational waves, we linearize the field equations around a flat Minkowski background $\eta_{\mu\nu}$

$$g_{\mu\nu} = \eta_{\mu\nu} + h_{\mu\nu} \qquad (2)$$

where $|h_{\mu\nu}| \ll 1$ is a small perturbation. In this approximation, the Einstein field equations can be simplified, resulting in the linearized equations:

$$\Box \bar{h}_{\mu\nu} = -\frac{16\pi G}{c^4} T_{\mu\nu} \qquad (3)$$

where $\Box = \eta^{\alpha\beta} \partial_\alpha \partial_\beta$ is the d'Alembertian operator, and $\bar{h}_{\mu\nu} = h_{\mu\nu} - \frac{1}{2}\eta_{\mu\nu}h$ is the trace-reversed perturbation.

## 2.3. Wave Equation in Vacuum:

In the vacuum where $T_{\mu\nu} = 0$, the wave equation simplifies to:

$$\bar{h}_{\mu\nu} = 0 \qquad (4)$$

This equation describes the propagation of gravitational waves in empty space.

## 2.4. Transverse-Traceless (TT) Gauge:

Gravitational waves can be further simplified by choosing the transverse-traceless (TT) gauge, where the perturbation $h_{\mu\nu}$ satisfies:

$$h_{0\nu} = 0, \quad \partial^i h_{ij} = 0, \quad h^i_i = 0 \qquad (5)$$

In this gauge, the nonzero components of $h_{\mu\nu}$ represent the physical degrees of freedom of the gravitational wave.

## 2.5. Plane Wave Solution:

A typical solution in the TT gauge for a plane gravitational wave traveling in the z-direction can be written as:

$$h^{TT}_{\mu\nu}(t,x) = \begin{pmatrix} 0 & 0 & 0 & 0 \\ 0 & h_+(t-z) & h_\times(t-z) & 0 \\ 0 & h_\times(t-z) & -h_+(t-z) & 0 \\ 0 & 0 & 0 & 0 \end{pmatrix} \qquad (6)$$

where $h_+$ and $h_\times$ are the plus and cross polarization states of the gravitational wave respectively.

## 2.6. Energy and Momentum:

The energy and momentum carried by gravitational waves can be described using the Isaacson stress-energy tensor in the limit of short wavelengths. The averaged energy $\langle T_{\mu\nu} \rangle$ for gravitational waves is given by:

$$\langle T_{\mu\nu} \rangle = \frac{c^4}{32\pi G} \langle \partial_\mu h_{\alpha\beta} \partial_\nu h^{\alpha\beta} \rangle \qquad (7)$$

In summary, gravitational waves are solutions to the linearized Einstein field equations in the vacuum, represented as transverse-traceless perturbations of the spacetime metric that propagate at the speed of light. Their mathematical description involves solving the wave equation for these perturbations and analyzing their effects and energy through the TT gauge and associated stress-energy tensor.

*2.7. Frequency spectrum*

The frequency spectrum of gravitational waves, which depends on the source, is composed mostly of low frequency events. While LIGO, an Earth based gravitational wave (GW) detection system has been successful in the direct detection of higher frequencies, and pulsar timing arrays have been successful in indirect hints of low frequencies, we are yet to have a GW detection system that directly makes observations in the low frequency range.

## 3. Significance of low frequency GW detection

Low frequency GWs, especially of the range 0.1 mHz to 0.1 Hz, are of significant interest as they arise from sources where the most physically exciting phenomena are hidden. These sources include [25][20][18]:

- Supermassive Black Hole Binaries (SBHBs)
- Intermediate-Mass Black Hole Binaries (IMBHBs)
- Extreme Mass Ratio Inspirals (EMRIs)
- Intermediate Mass Ratio Inspirals (IMRIs)
- Stellar Origin Black Hole Binaries (SOBHBs)
- Galactic Binaries
- Cosmological sources that produce a stochastic background

Analysing GWs from these sources may help drive research in fundamental physics ahead by delivering us significant clues regarding the nature of black holes, SEP and screening, Dark Energy and screening, dark matter and primordial black holes, astrophysical systematics and more.

## 4. Laser Interferometer Space Antenna (LISA)

LISA is a space-based gravitational wave observatory developed by the European Space Agency (ESA) with contributions from NASA. [15] It is designed to detect gravitational waves in the frequency range of 0.1 mHz to 0.1 Hz, which are not accessible to ground-based detectors due to seismic, thermal, and gravity gradient noise. The components of LISA include [23][14][13][8][3]:

*4.1. Configuration and structure*

LISA consists of three spacecraft (SC1, SC2, SC3) arranged in an equilateral triangle formation with an arm length of $2.5×10^6$ km. The spacecraft are in a heliocentric orbit trailing Earth by approximately 20°.

   a. Test masses (TM):
      - Material: Platinum/Gold alloy
      - Dimensions: 46 mm by 46 mm by 46 mm
      - Mass: 1.96 kg
      - Each spacecraft contains two test masses housed in a gravitational reference sensor (GRS) to minimise non-gravitational forces.
   b. Optical Bench (OB):
      - Made from Zerodur/ULE glass to minimise thermal expansion.

- Houses interferometric optics to combine and measure laser beams.

*4.2. Laser Interferometry*

LISA uses heterodyne laser interferometry to measure changes in the distance between the test masses. Each spacecraft sends and receives laser beams to and from the other two spacecraft, forming multiple Michelson-type interferometers.

a. Laser System:
   - Laser Type: Nd:YAG laser.
   - Wavelength: 1064 nm.
   - Output Power: 2 W.
   - Frequency Stability: $\Delta v/v \approx 10^{-21}$ over the mission duration.

b. Measurement Principle:
   - The interferometric measurement involves the transmission of laser beams between spacecraft, with phase shifts indicating changes in distance.
   - Laser phase noise is reduced using the Time Delay Interferometry (TDI) technique, which synthesizes multiple time-delayed signals to cancel common-mode noise.

Interferometry is of crucial importance because the phase shifts in the fringe provide us with significant information about the nature of gravitational waves. It is also worth noting that a slight deviation in the length of the triangle formed by LISA spacecrafts will completely nullify the fringe formation. Another important factor that also disturbs the fringe formation is phase noise, which we have studied in depth in this project.

*4.3. Orbits and Formation Flying*

The spacecraft follow heliocentric orbits, maintaining a triangular formation. The arm lengths vary slightly due to Keplerian motion, but the relative positions are controlled to within a few tens of meters.

a. Formation Control:
   - Micropropulsion systems, such as colloidal thrusters, are used to maintain the formation.
   - Autonomous control algorithms adjust the spacecraft positions to ensure the test masses remain centered within their housings.
b. Orbital Parameters:
   - Semi-major axis: 1 AU.
   - Eccentricity: Low, to maintain near-circular orbits.

*4.4. Noise Reduction*

LISA employs several techniques to minimize noise and ensure high-precision measurements.

a. Drag-Free Control:
   - Each spacecraft uses a drag-free control system where the spacecraft follows the test mass, effectively shielding it from non-gravitational forces.
   - Micro-thrusters counteract perturbations from solar radiation pressure and other disturbances.
b. Thermal and Radiation Shielding:

- The test masses and optical components are housed in a thermally controlled environment to reduce thermal noise.
- Shielding against cosmic rays and solar particles is provided to minimize radiation noise.

*4.5. Data Analysis and Signal Processing*

LISA's data analysis involves extracting gravitational wave signals from the raw interferometric measurements, which are transmitted to Earth for further processing.

a. Time Delay Interferometry (TDI):
   - A technique to cancel out laser frequency noise by creating virtual equal-arm interferometers using time-delayed signals.
   - TDI accounts for the varying distances between spacecraft and ensures high sensitivity to gravitational waves.
b. Source Localization:
   - By combining signals from the three spacecraft, LISA can determine the direction and polarization of incoming gravitational waves.
   - Algorithms reconstruct the waveforms and extract astrophysical parameters of the sources.

*4.6. Technical Specifications*

a. Spacecraft Dimensions:
   - Each spacecraft is approximately 3 m in diameter with solar arrays extending up to 10 m.
b. Power System:
   - Solar arrays generate power for onboard systems and instruments.
   - Power consumption per spacecraft is approximately 500 W.
c. Communication System:
   - High-gain antennas provide data transmission to Earth.
   - Data rate: Up to 1 Mbps.

It is worth noting that the telescopes mentioned here will be subject to wavefront aberrations. So, before we discuss the details, we will have a quick overview of wavefront aberrations.

## 5. Wavefront Aberrations

Wavefront aberrations in an optical system can be mathematically represented as deviations from an ideal wavefront. These aberrations are typically described using Zernike polynomials, which provide a complete set of orthogonal functions over a unit circle. The aberrations are expressed in terms of these polynomials, each corresponding to a particular type of wavefront distortion. [16][17]

*5.1. Wavefront Aberration Representation*

Let $\Phi(\rho,\theta)$ represent the wavefront aberration function in polar coordinates $(\rho,\theta)$, where $\rho$ is the radial coordinate (normalized to the unit circle) and $\theta$ is the azimuthal angle. The aberration function $\Phi(\rho,\theta)$ can be expanded as a series of Zernike polynomials:

$$\Phi(\rho,\theta) = \sum_{n=0}^{\infty} \sum_{m=-n}^{n} c_{nm} Z_{nm}(\rho, \theta) \qquad (8)$$

Here, $c_{nm}$ are the coefficients corresponding to the Zernike polynomials $Z_{nm}(\rho,\theta)$. These coefficients quantify the contribution of each aberration term.

## 5.2. Zernike Polynomials

The Zernike polynomials $Z_{nm}(\rho,\theta)$ defined as:

$$Z_{nm}(\rho,\theta) = R_n^m(\rho) \begin{cases} \cos(m\theta) & \text{if } m \geq 0 \\ \sin(|m|\theta) & \text{if } m < 0 \end{cases} \quad (9)$$

where $R_n^m(\rho)$ are the radial polynomials given by:

$$R_n^m(\rho) = \sum_{s=0}^{(n-m)/2} \frac{(-1)^s (n-s)!}{s!\left(\frac{n+m}{2}-s\right)!\left(\frac{n-m}{2}-s\right)!} \rho^{n-2s} \quad (10)$$

## 5.3. Types of Aberrations

The different types of wavefront aberrations are associated with specific Zernike polynomials and can be classified as follows:

a. Rotationally Symmetric Aberrations (RSAs)

RSAs are aberrations that do not depend on the azimuthal angle $\theta$ and are symmetric around the optical axis. These aberrations can be represented using Zernike polynomials where $m=0$. This symmetry means that the wavefront distortion is the same at any angle around the axis of symmetry. Common RSAs include:

- Piston ($Z_0^0$):

$$Z_0^0(\rho) = 1 \quad (11)$$

Represents a constant phase shift across the entire wavefront.

- Defocus ($Z_2^0$):

$$Z_2^0(\rho) = 2\rho^2 - 1 \quad (12)$$

Represents a quadratic phase change, symmetric around the optical axis.

- Spherical Aberration ($Z_4^0$):

$$Z_4^0(\rho) = 6\rho^4 - 6\rho^2 + 1 \quad (13)$$

Represents a higher-order radial phase change, also symmetric around the optical axis.

The general form of RSAs can be expressed as:

$$\Phi_{RSA}(\rho) = \sum_{n \text{ even}} c_{n0} R_n^0(\rho) \quad (14)$$

b. Non-Rotationally Symmetric Aberrations (NRSAs)

NRSAs are aberrations that vary with the azimuthal angle $\theta$ and are not symmetric around the optical axis. These aberrations are represented using Zernike polynomials where $m \neq 0$. The wavefront distortion introduced by NRSAs changes with the azimuthal angle, leading to asymmetrical patterns. Common NRSAs include:

- Tilt ($Z_1^1$ and $Z_1^{-1}$):

$$Z_1^1(\rho,\theta) = \rho\cos(\theta), \quad Z_1^{-1}(\rho,\theta) = \rho\sin(\theta) \quad (15)$$

Represent linear phase changes in orthogonal directions, dependent on $\theta$.

- Astigmatism ($Z_2^2$ and $Z_2^{-2}$):

$$Z_2^2(\rho,\theta)=\rho^2\cos(2\theta), \qquad Z_2^{-2}(\rho,\theta)=\rho^2\sin(2\theta) \qquad (16)$$

Represent cylindrical distortions that vary with twice the azimuthal angle.

- Coma ($Z_3^1$ and $Z_3^{-1}$):

$$Z_3^1(\rho,\theta)=(3\rho^3-2\rho)\cos(\theta), \qquad Z_3^{-1}(\rho,\theta)=(3\rho^3-2\rho)\sin(\theta) \qquad (17)$$

Represent asymmetric distortions that introduce a comet-like tail.

The general form of NRSAs can be expressed as:

$$\Phi_{\text{NRSA}}(\rho,\theta)=\sum_{n\ even}\sum_{\neq 0}c_{nm}Z_{nm}(\rho,\theta) \qquad (18)$$

*5.4. Practical Implications in Optical Systems*

Wavefront aberrations introduce phase errors in the optical system, which can be detrimental to applications requiring high precision, such as gravitational wave detection. The phase error $\Delta\varphi(\rho,\theta)$ introduced by an aberrated wavefront can be expressed as:

$$\Delta\varphi(\rho,\theta) = \frac{2\pi}{\lambda}\Phi(\rho,\theta) \qquad (19)$$

where $\lambda$ is the wavelength of the light. The goal in designing optical systems is to minimize these aberrations to achieve the required level of precision.

## 6. Sources of noise on LISA

Apart from aberrations, the accuracy of readings of LISA telescopes will also be subject to various kinds of noise. Some of the prominent sources of noise are likely to be the following:

- Shot Noise**:** This arises from the quantization of light and is a fundamental limit in any optical measurement. It depends on the intensity of the light and the efficiency of the photodetectors. In LISA, shot noise is expected to be a significant source of noise, especially at higher frequencies.
- Acceleration Noise: LISA will experience various accelerations due to non-gravitational forces such as solar radiation pressure, spacecraft thruster noise, and residual gas drag. These accelerations can induce noise in the interferometric measurements.
- Phase Noise**:** Phase noise can arise from various sources such as laser frequency instability, fluctuations in the optical path length, and thermal effects. Minimizing phase noise is crucial for maintaining the stability of the interferometric measurements.
- Instrumental Noise**:** This includes noise from the electronics, readout systems, and other components of the instrument. Careful design and calibration are necessary to minimize instrumental noise contributions.
- Environmental Noise: Cosmic rays, micrometeoroid impacts, and other environmental factors can introduce noise into the system. Shielding and redundancy measures may be employed to mitigate these effects.
- Data Processing Noise**:** Noise can also be introduced during data processing and analysis. Algorithms for signal extraction and parameter estimation should be robust against various sources of noise.

We will focus on the significance of phase noise further in this project, where it leads to results that are of direct importance in the development of LISA.

## 7. Wavefront simulation with all aberrations present

The analysis was carried out by creating a python code (packaged with matplotlib) that accounts for aberrations on a simulated optical design of the LISA telescope. The wavelength is equal to 1064 nm, aperture radius is 0.15 metres, and the grid size is 500. After successfully running the simulation, we obtain the following results:

- Simulated wavefront phase aberration (inclusive of all aberration types)

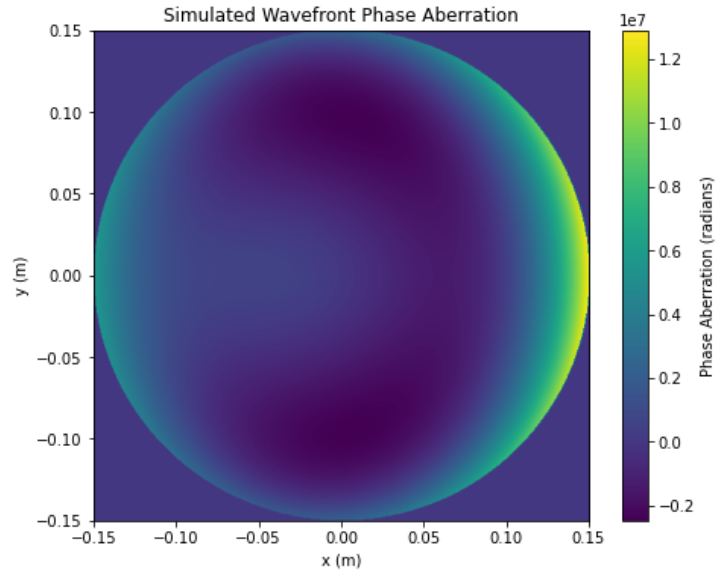

*Figure 1*

- Piston

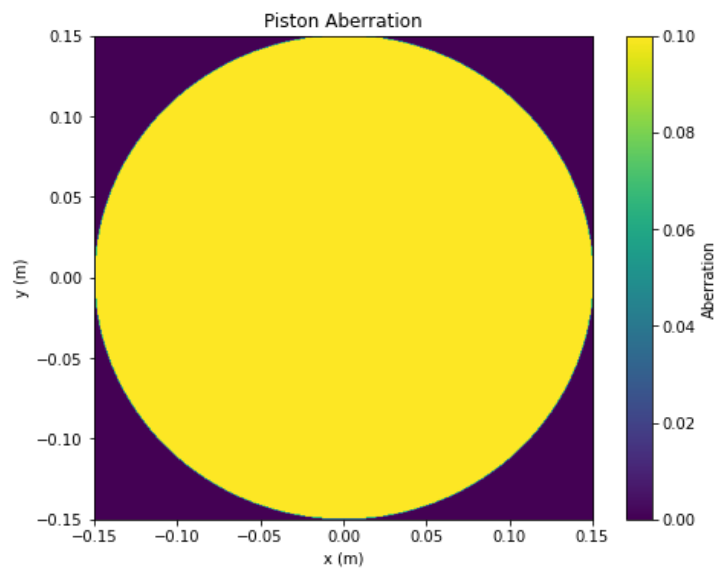

*Figure 2*

- Tilt x

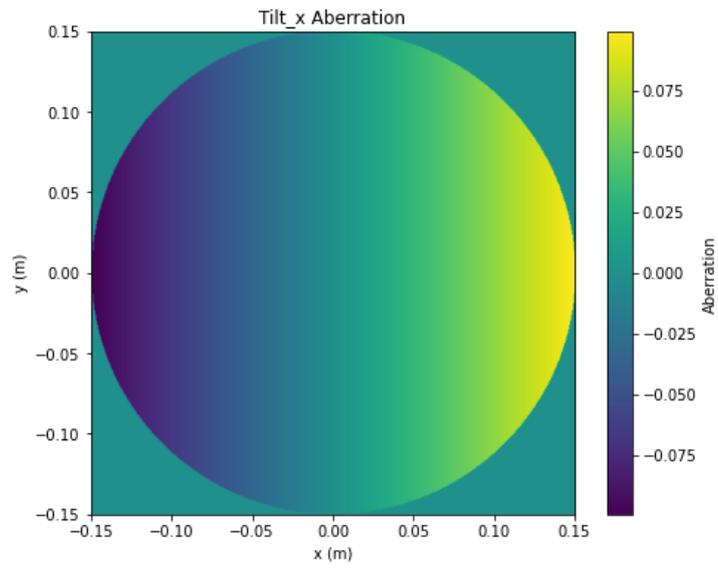

*Figure 3*

- Tilt y

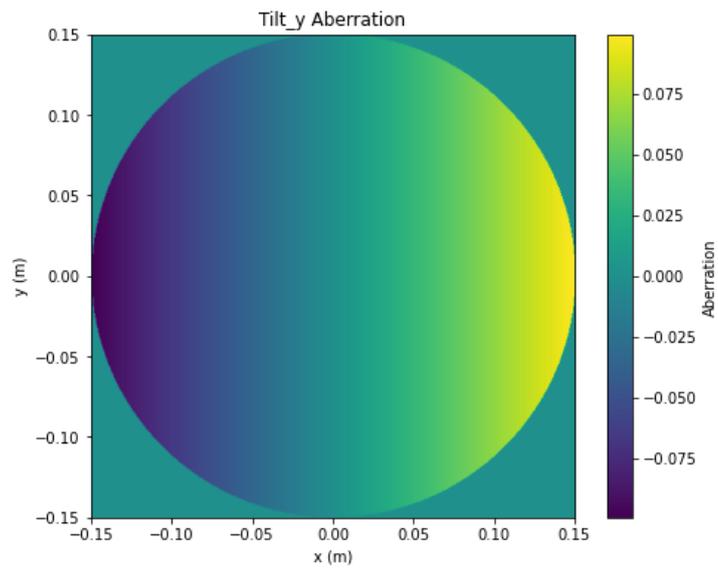

*Figure 4*

- Defocus

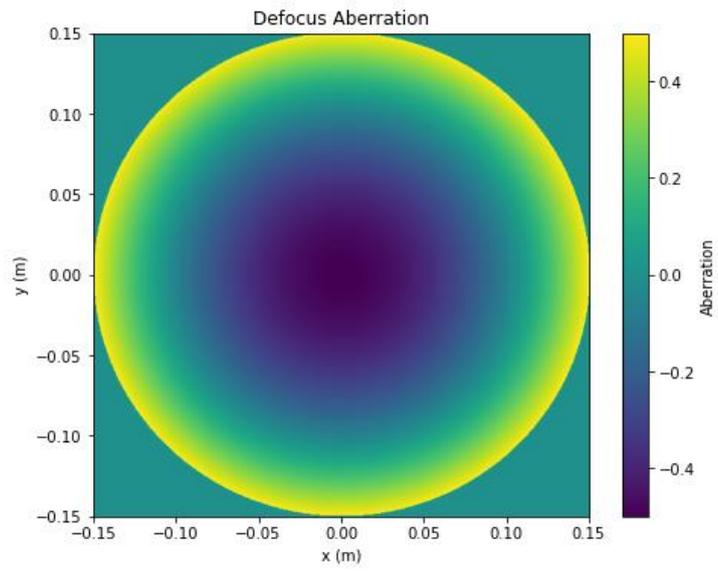

*Figure 5*

- Astigmatism 0 degree

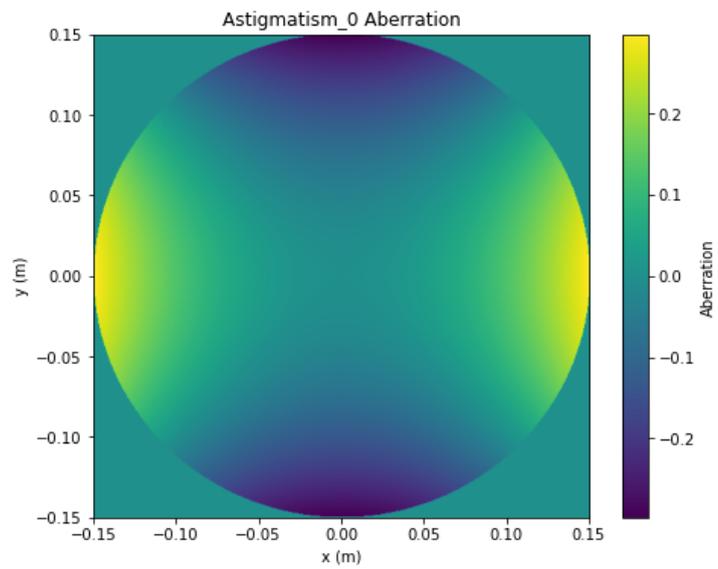

*Figure 6*

- Astigmatism 45 degree

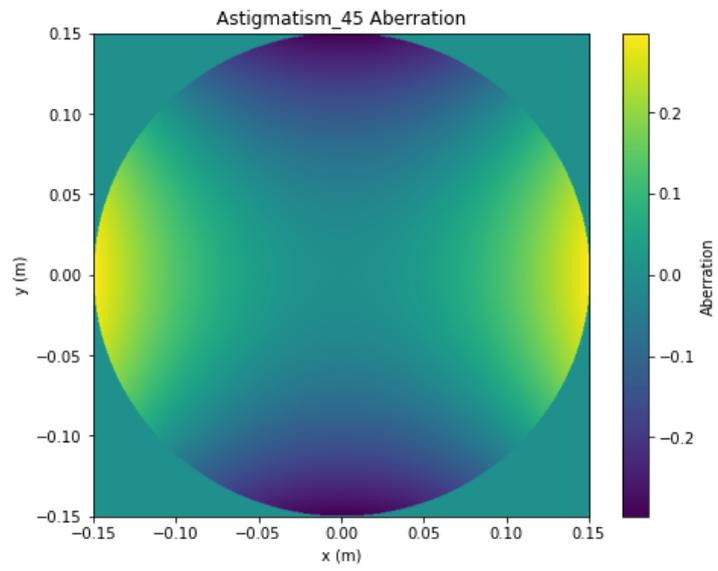

*Figure 7*

- Coma x

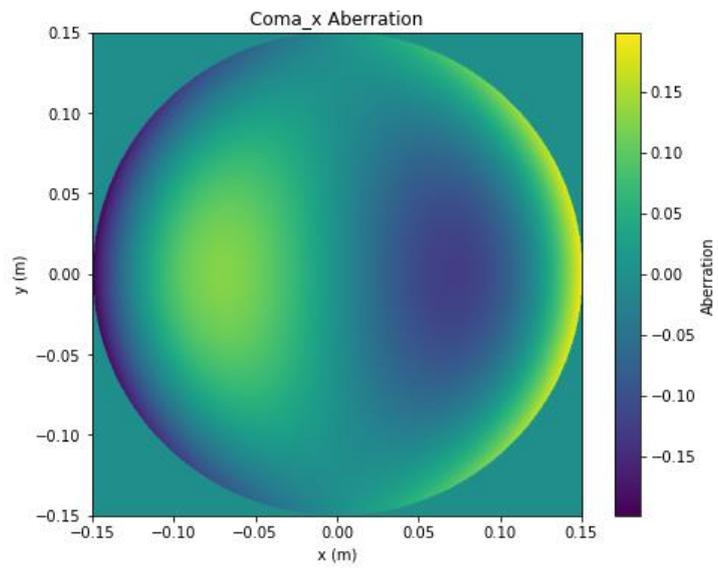

*Figure 8*

- Coma y

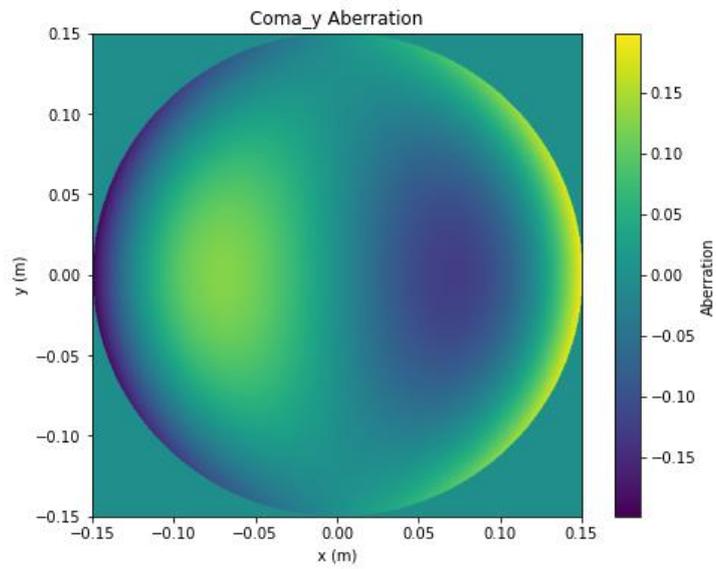

*Figure 9*

- Spherical aberration

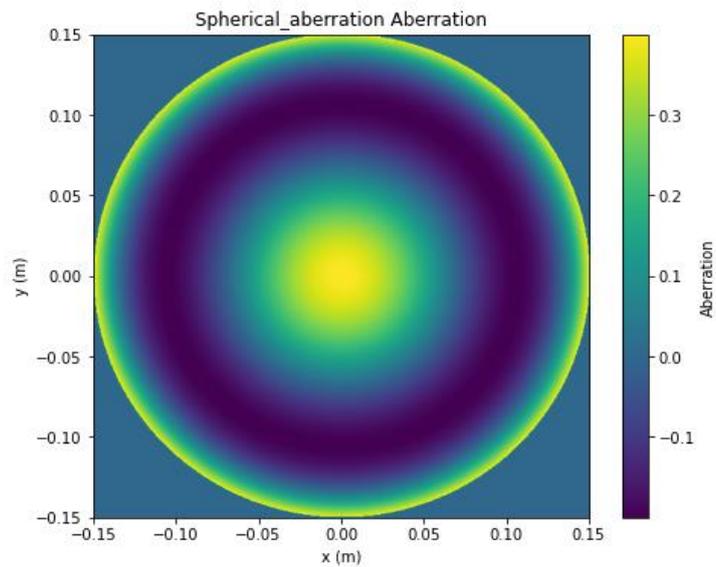

*Figure 10*

- Interference pattern without noise (intensity plotted here is relative, based on the difference in intensities of different regions in the plot)

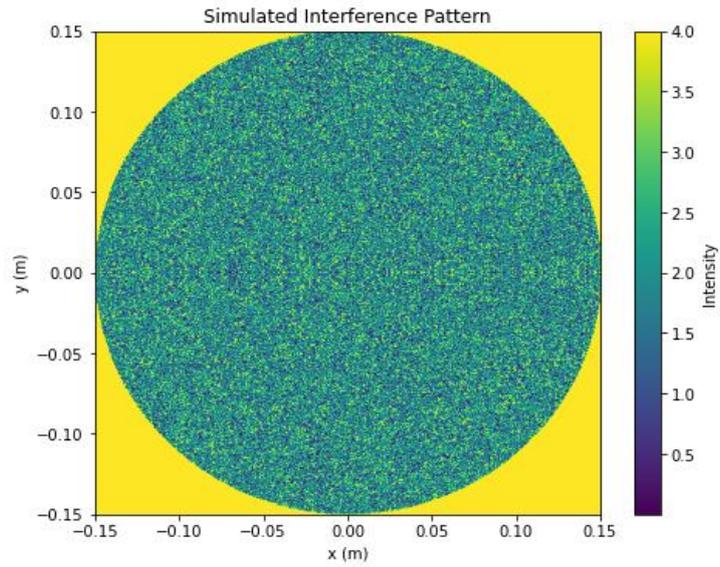

*Figure 11*

- Interference pattern with noise = 1 nm (RMS value)

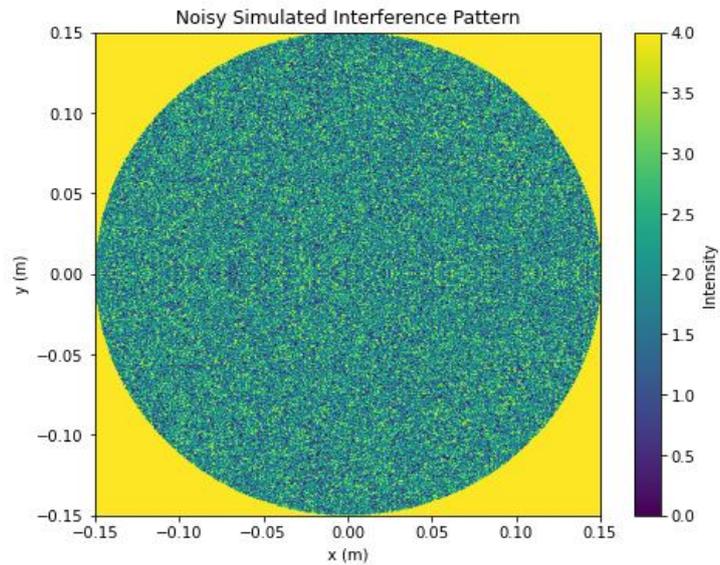

*Figure 12*

It is evident from the results above that aberrations on the primary mirror in LISA are high primarily due to rotationally symmetric aberrations. Several optimisation techniques have been implemented to account for most of them, while neglecting rotationally asymmetric aberration types. [4][2][26]

Consequently, in this work, we will also focus on the most neglected aberration types, i.e. defocus and astigmatism, and deduct whether they are worthy of consideration in the development of LISA.

## 8. Wavefront simulation after optimisation

After correcting piston, defocus, spherical, and coma aberration types in the simulation, as commonly done by several papers, we are often left with tilt and astigmatism aberration types in the system. It has been well verified by several authors that the two of these aberration types do not have much effect on the system and it is not a necessity to correct them for a good functioning of LISA. We will now try to verify the same (aberration plotted in radians):

- Total aberration

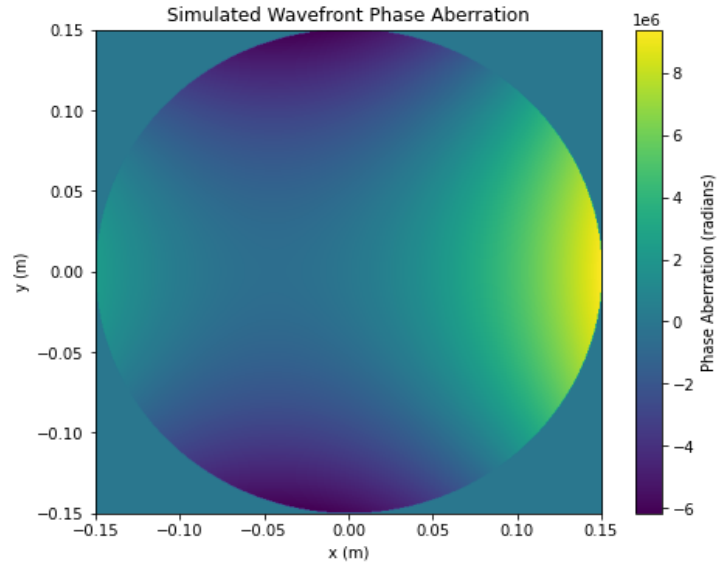

*Figure 13*

- Tilt x

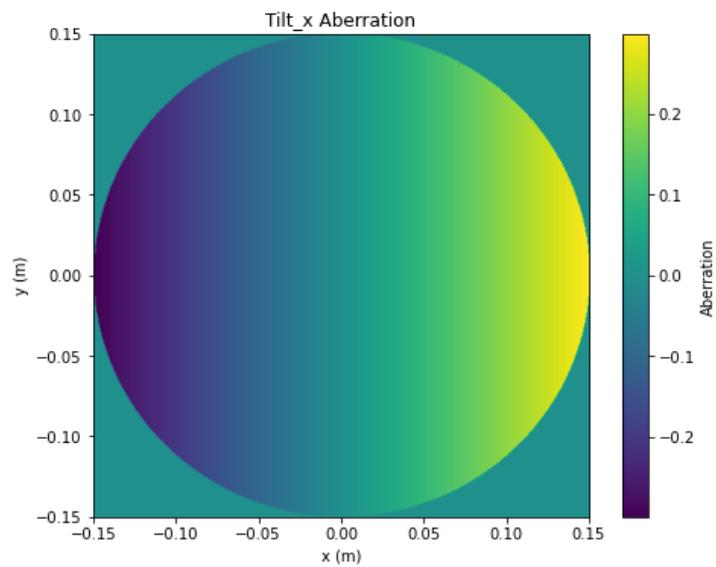

*Figure 14*

- Tilt y

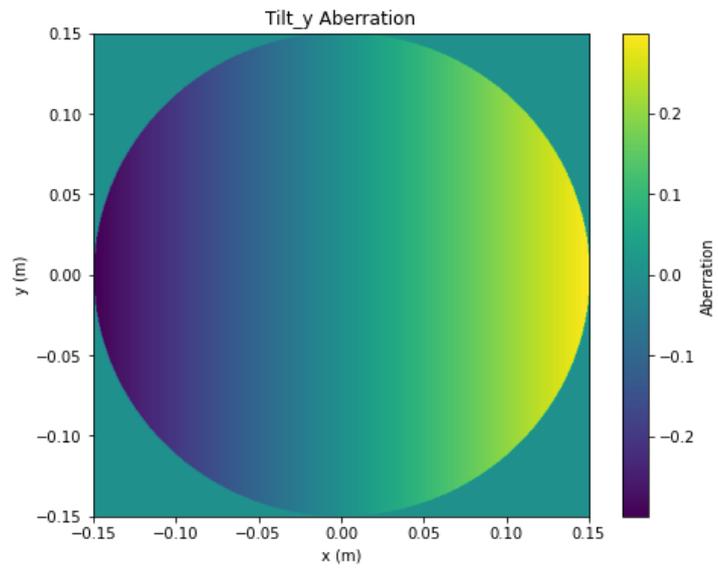

*Figure 15*

- Astigmatism 0 degree

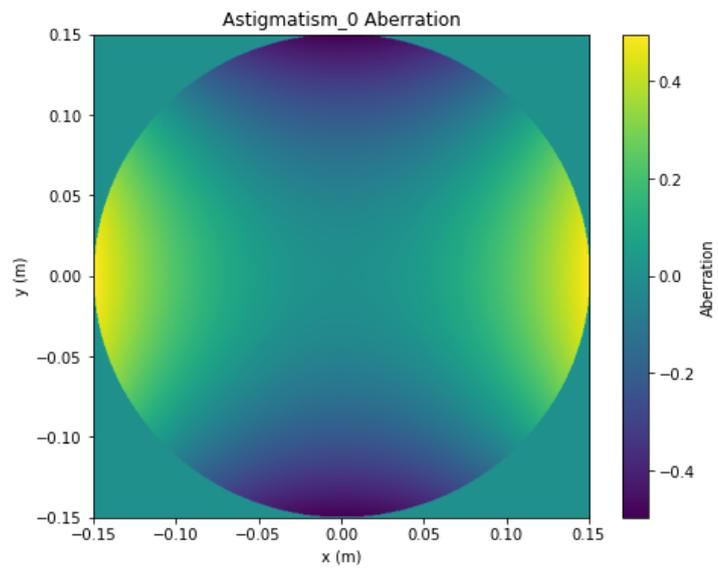

*Figure 16*

- Astigmatism 45 degrees

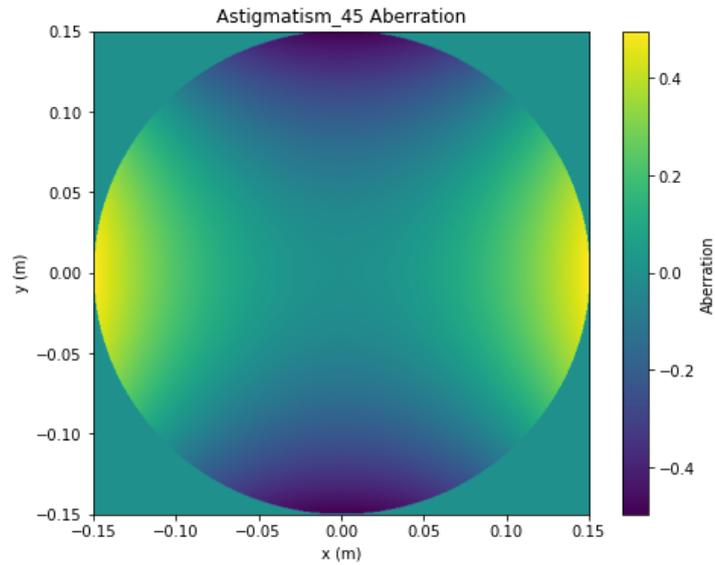

*Figure 17*

- Interference fringe without noise

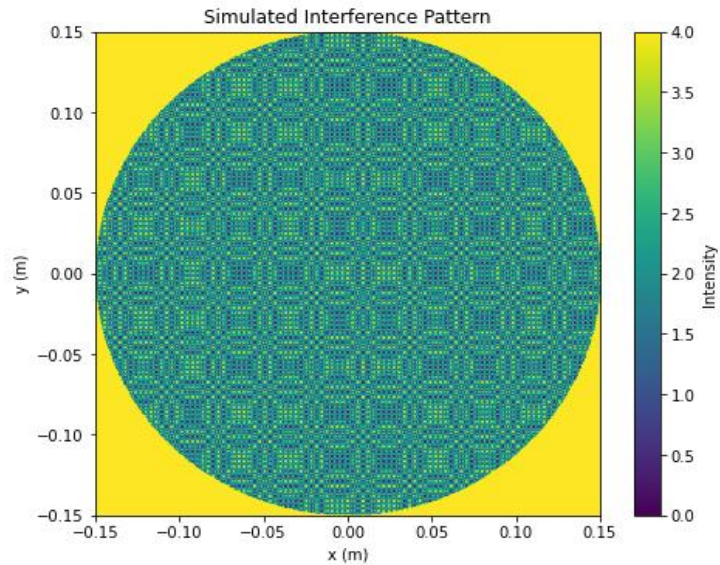

*Figure 18*

The fringe visibility, as it turns out, decreases with increase in phase noise in the system. The following is the curve of fringe visibility with increase in noise in the system. As visible in figure 19, the curve is close to a logarithmic decay. The reasoning behind the slowing down of fringe visibility difference is that the fringe visibility is approaching zero after roughly 150 nm of noise.

The equation we get for the fitted curve is:

$$y = 6.242 \times 10^{-8} x^3 - 1.889 \times 10^{-5} x^2 + 5.802 \times 10^{-4} x^{-0.005405} \tag{20}$$

where y is the fringe visibility difference (where 0 corresponds to no drop in fringe visibility, i.e. perfect fringe visibility) and x is phase noise.

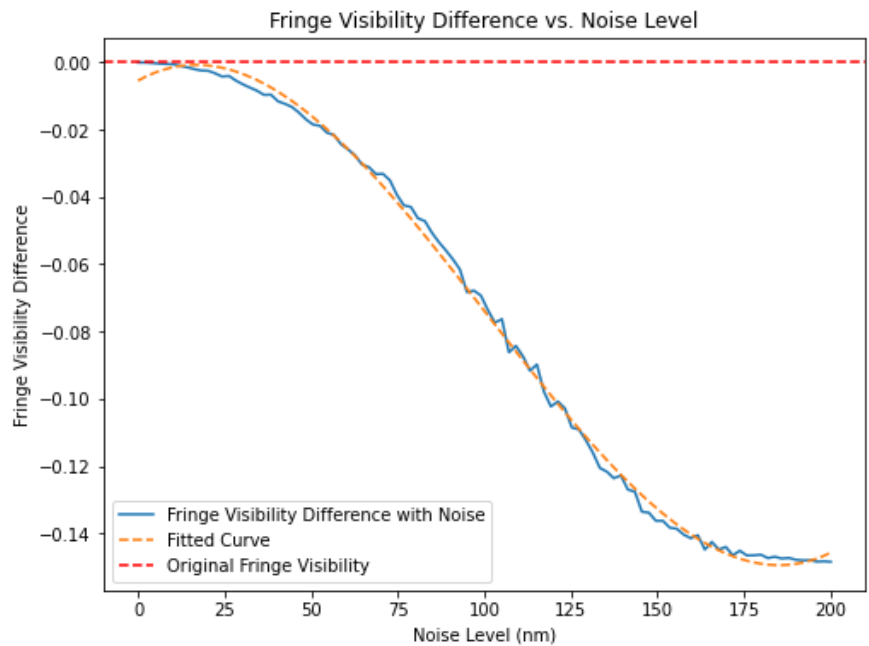

*Figure 19*

## 9. Analysis of the Observation

The observation that the interference fringe visibility approaches zero when the noise value is above 150 nm indicates the sensitivity of the interferometric system to phase noise. This sensitivity is crucial for understanding the system's performance and the impact of various noise levels on the measurement accuracy.

### 9.1. Interferometric Sensitivity

Interferometric systems like LISA are extremely sensitive to phase changes, which can be caused by a variety of factors, including thermal fluctuations, mechanical vibrations, and other environmental disturbances. When the phase noise level reaches a certain threshold, it can significantly degrade the quality of the interference pattern, making it difficult to distinguish the signal from the noise.

### 9.2. Phase Noise Impact

The phase aberration introduced by the wavefront distortions and the added phase noise are both measured in radians. The relationship between the phase noise and the interference pattern's visibility can be understood by considering how the phase affects the constructive and destructive interference:

   a. Constructive Interference: Occurs when the phase difference between the reference and the measured wavefronts is an integer multiple of $2\pi$, resulting in maximum intensity.
   b. Destructive Interference: Occurs when the phase difference is an odd multiple of $\pi$, resulting in minimum intensity.

When the phase noise becomes significant (150 nm), it generates such high random variations in the phase differences across the wavefront, that they disrupt the constructive and destructive interference patterns.

*9.3. Quantitative Understanding*

To quantify this, consider the phase shift introduced by a noise level of 150 nm:

$$\Delta\phi = \frac{2\pi \times 150 \times 10^{-9}}{1064 \times 10^{-9}} \approx 0.886 \text{ radians} \qquad (21)$$

A phase shift of approximately 0.886 radians is significant because it represents a considerable fraction of a wavelength. This level of phase noise can cause substantial distortions in the interference pattern, making it difficult to maintain the precise phase relationships needed for accurate measurements.

*9.4. Implications*

a. System Design and Noise Mitigation: This observation underscores the importance of noise mitigation strategies in the design of interferometric systems like LISA. Techniques such as thermal shielding, vibration isolation, and precision manufacturing are critical to minimizing phase noise.
b. Signal Processing: Advanced signal processing techniques may be required to extract the signal from the noisy interference pattern. This could involve filtering, phase unwrapping, and other methods to enhance the signal-to-noise ratio.
c. Performance Requirements: Understanding the noise threshold at which the interference pattern loses visibility helps set performance requirements for the system. For LISA, maintaining phase noise well below 150 nm is essential to achieve the desired measurement accuracy.

## 10. Conclusion

In this report, we studied the physics of gravitational waves, their detection techniques, and the implications of low frequency GW detection for the advancement of fundamental physics.

We also studied the structure and functioning of LISA, while giving special emphasis on its optical and interferometric systems.

Then, we analyzed the wavefront aberrations on a simulation of LISA primary telescope using a python program which agrees with the work of other authors, highlighting the importance of optimising rotationally symmetric aberrations while neglecting rotationally asymmetric aberrations.

Finally, we carried out some research to verify whether the negligence of rotationally asymmetric aberrations can create issues, and we found out that rotationally asymmetric aberrations, especially tilt and astigmatism, can lead to the loss of fringe visibility, an essential component of GW detection, when noise levels are higher than a certain threshold (150 nm). This is because fringe visibility and noise have an inverse relationship.

This result can have significant implications in the design and development of LISA telescopes as it gives an upper threshold of allowed noise (150 nm) in the system provided tilt and astigmatism aberrations are left remaining in the system.

The fringe visibility loss observed in the interference pattern at a noise level of 150 nm and higher highlights the sensitivity of the LISA interferometric system to phase noise. This observation informs the need for stringent noise control measures and advanced signal processing techniques to ensure accurate and reliable measurements. Maintaining phase stability at very low levels is crucial for the success of high-precision interferometric missions like LISA.

## 11. Acknowledgement

I sincerely thank Dr. Marco De Petris for teaching me the optics of astronomical systems in the course 'Astronomical Optics' at Sapienza University of Rome, which helped bring this project to life.